\documentclass[reprint,twocolumn,aps,superscriptaddress]{revtex4-1}
\usepackage[utf8]{inputenc}
\usepackage[T1]{fontenc}
\usepackage{amsmath}
\usepackage{amsfonts}
\usepackage{physics}
\usepackage[caption=false]{subfig}
\usepackage{bm}
\usepackage{amssymb}
\usepackage[mathscr]{eucal}
\usepackage{graphicx}
\usepackage[dvipsnames]{xcolor}
\usepackage[hidelinks,colorlinks=true,allcolors=Blue]{hyperref}

\begin{document}
\title{Blackbody radiation and thermal effects on chemical reactions and phase transitions in cavities}
\author{Sindhana Pannir-Sivajothi}
\affiliation{Department of Chemistry and Biochemistry, University of California
San Diego, La Jolla, California 92093, USA}
\author{Joel Yuen-Zhou}
\email{joelyuen@ucsd.edu}

\affiliation{Department of Chemistry and Biochemistry, University of California
San Diego, La Jolla, California 92093, USA}
\begin{abstract}
An important question in polariton chemistry is whether reacting molecules are in thermal equilibrium with their surroundings. If not, can experimental changes observed in reaction rates of molecules in a cavity (even without optical pumping) be attributed to a higher/lower temperature inside the cavity? In this work, we address this question by computing temperature differences between reacting molecules inside a cavity and the air outside. We find this temperature difference to be negligible for most reactions. On the other hand, for phase transitions inside cavities, as the temperature of the material is actively maintained by a heating/cooling source in experiments, we show cavities can modify observed transition temperatures when mirrors and cavity windows are ideal (non-absorbing); however, this modification vanishes when real mirrors and windows are used. Finally, we find substantial differences in blackbody spectral energy density between free space and infrared cavities, which reveal resonance effects and could potentially play a role in explaining changes in chemical reactivity in the dark.
\end{abstract}
\maketitle

 When a material is placed inside a Fabry-Perot cavity, experiments report changes in chemical reaction rates \cite{thomas2016ground,thomas2019tilting,ahn2023modification,vergauwe2019modification,hirai2020modulation,lather2020improving,lather2019cavity} and phase transition temperatures \cite{wang2014phase,brawley2023sub, jarc2023cavity} even in the absence of optical pumping. Several theoretical explanations have been proposed for these cavity-mediated effects \cite{galego2015cavity,herrera2016cavity,flick2018cavity,campos2019resonant,li2021cavity,schafer2022shining}, but there is no consensus on what causes changes in reactivity \cite{fregoni2022theoretical,mandal2023theoretical,campos2023swinging,ruggenthaler2023understanding,herrera2020molecular,sidler2022perspective}. They are often attributed to polariton modes that emerge under strong light-matter coupling; however, most of these experiments are performed in the collective strong coupling regime where theory predicts negligible modification of reaction rates \cite{vurgaftman2020negligible,campos2020polaritonic,zhdanov2020vacuum,li2020origin} and phase transitions \cite{pilar2020thermodynamics}. In this regime, a large number of molecular modes, $N$, collectively couple to a single photon mode to form two polariton and $N-1$ dark modes. Here, the collective coupling strength surpasses losses from the system. The dark modes, devoid of any photon character and expected to behave similarly to bare molecular modes, outnumber the hybrid polariton modes by a factor of $N\sim10^8-10^{11}$ and dominate the kinetics and thermodynamics of the system; disorder does not qualitatively change the conclusion \cite{du2022catalysis}.
 
 The disparity between theory and experiment prompts a critical examination of whether assumptions in theoretical models align with conditions present in the experimental setup. In this work, we specifically investigate a crucial assumption in polariton chemistry: that the temperature of the material within the cavity matches that of the surrounding air. This is assumed despite the fact that heat is generated/absorbed during a reaction. Given that cavities can modify the rate of heat transfer—an aspect applied even in practical scenarios such as radiative cooling of photovoltaic cells \cite{taylor2017vanadium,cho2023directional}—this question becomes both important and reasonable. We proceed to quantify whether changes in experimentally observed reaction rates and phase transition temperatures can be attributed to temperature differences between the material within the cavity and external air. These inquiries are in line with recent experiments that emphasize the role of non-polaritonic effects in cavity systems \cite{thomas2023non}. Furthermore, we also compute differences between blackbody radiation inside a cavity and in free space, which could play a role in explaining the aforementioned experiments. 

\begin{figure}
\includegraphics[width=\columnwidth]{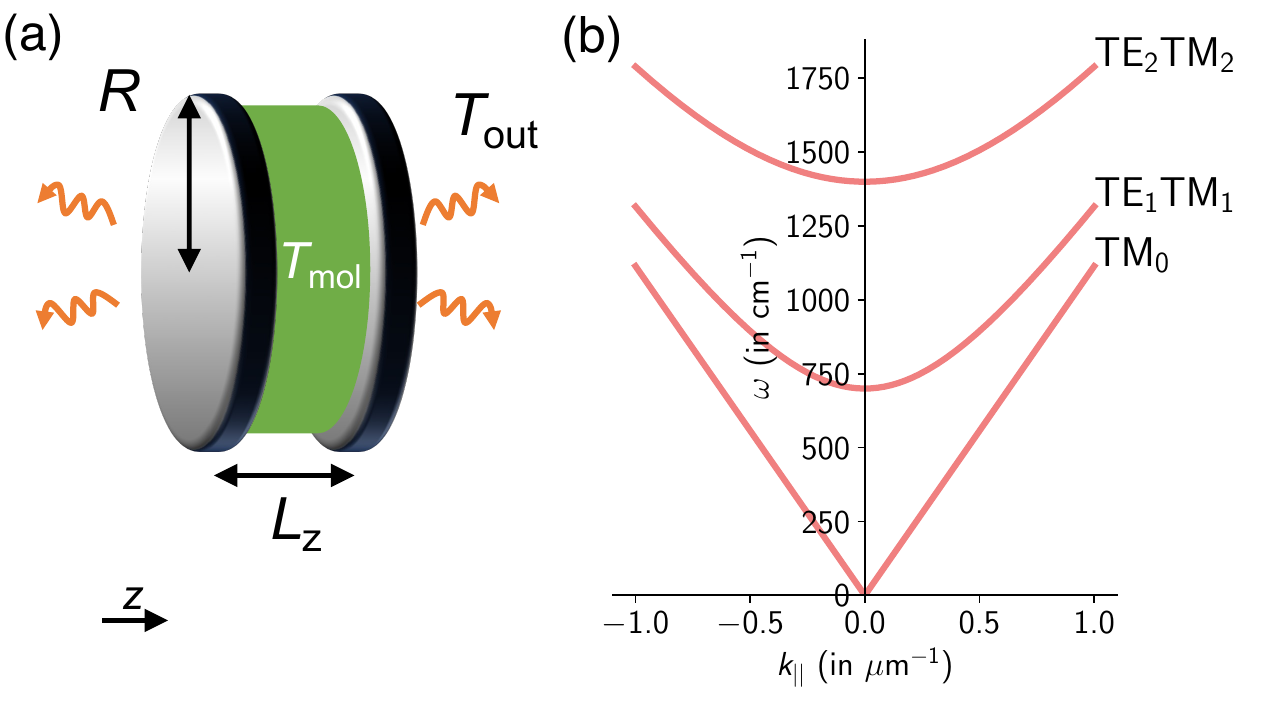}
	\caption{\label{fig:1} \textbf{Fabry-Perot cavity modes.} (a) Cavity with molecules at temperature $T_{\mathrm{mol}}$ different from that of external air $T_{\mathrm{out}}$. (b) Modes of a cavity with fundamental frequency $\omega_{\mathrm{cav}}=700\mathrm{cm}^{-1}$.}
\end{figure}

\section*{Results and discussion}

We study heat transfer from a Fabry-Perot cavity to the air surrounding it (Fig. \ref{fig:1}). In our model, we include all modes of a Fabry-Perot cavity as multimode models capture important qualitative features missed by single-mode ones
\cite{ribeiro2022multimode,ying2023resonance,engelhardt2023polariton}. The dispersion relation of photon modes within such a cavity, depicted in Fig. \ref{fig:1}b, is given by:
\begin{equation}
	\omega_{m}(k_{||})=\frac{c}{n}\sqrt{k_{||}^2+\Big(\frac{m\pi}{L_z}\Big)^2}
\end{equation}
when the mirrors are ideal (100\% reflectivity) \cite{ellingson2020electromagneticsv2}.
Here, $m$ labels the branch of the cavity modes, $k_{||}$ is the wavevector component parallel to the mirrors, $n$ the refractive index of the material filling the cavity, $c$ the speed of light, and $L_z$ the distance between the mirrors. For each $m>0$, two polarizations of modes exist: TE and TM. When $m=0$, only TM$_0$ modes exist which are not observed in the reflection spectrum in experiments due to zero $z$ component of their wavevector $\mathbf{k}$. Consequently, the lowest observed frequency, corresponding to $m=1$ at normal incidence ($k_{||}=0$), is labeled the cavity frequency $\omega_{\mathrm{cav}}=c\pi/nL_z$.

\begin{figure}
\includegraphics[width=0.75\columnwidth]{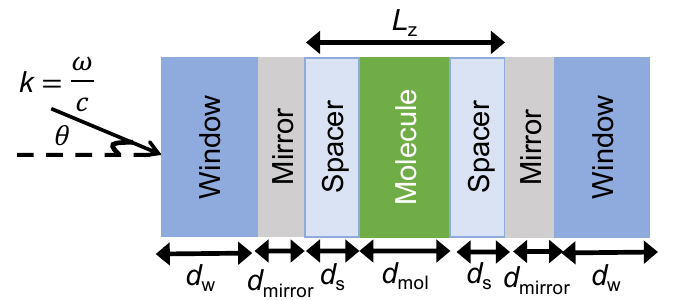}
	\caption{\label{fig:transfer} \textbf{Transfer matrix simulation.} The structure used for transfer matrix simulations to calculate the spectral directional emissivity $\mathcal{E}_h(\mathbf{k})$ where $h=\mathrm{TE,TM}$. It contains windows, mirrors, spacers and molecules of thicknesses $d_{\mathrm{w}}$, $d_{\mathrm{mirror}}$, $d_{\mathrm{s}}$, and $d_{\mathrm{mol}}$, respectively.}
\end{figure}

\begin{figure*}
	\includegraphics[width=\textwidth]{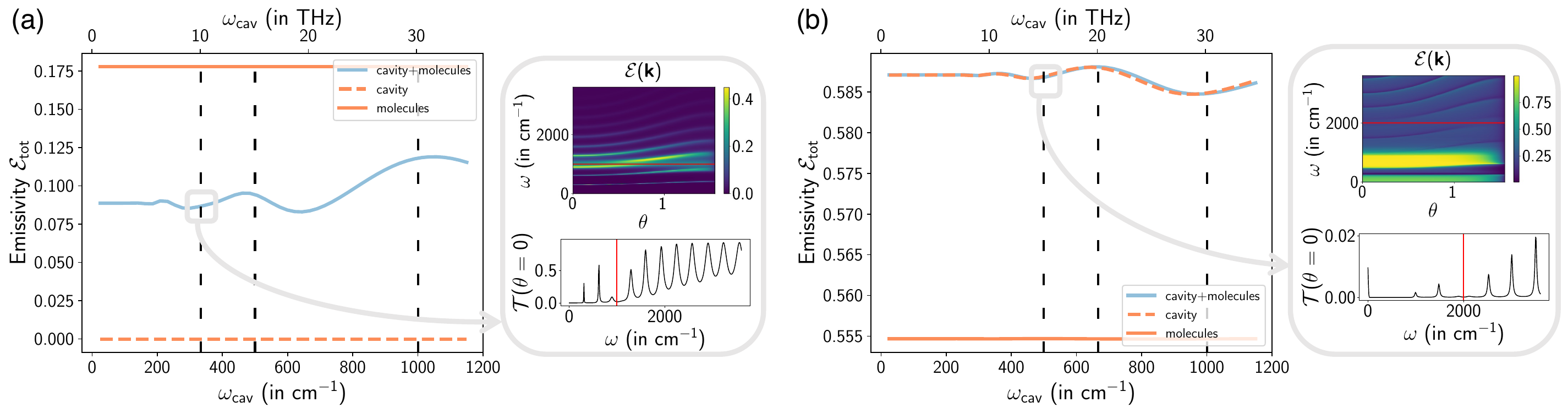}
	\caption{\label{fig:emissivity} \textbf{Emissivity.} The emissivity $\mathcal{E}_{\mathrm{tot}}$ is plotted against cavity frequency $\omega_{\mathrm{cav}}=c\pi/nL_z$ for cavity+molecules (blue line), molecules (solid orange line), and cavity without molecules (dashed orange line). Two systems were considered: (a) an ideal system with non-absorbing mirrors and windows with the molecular absorption centered at $\omega_{\mathrm{mol}}=1000\mathrm{cm}^{-1}$ and (b) a realistic system with absorbing mirrors and windows with the molecular absorption centered at $\omega_{\mathrm{mol}}=2000\mathrm{cm}^{-1}$. In (a), we use real-valued refractive indices for the spacer, window, and mirror while in (b), we use complex-valued refractive indices of Au for mirrors, CaF$_2$ for windows, and $n_{\mathrm{s}}=1.43$ for spacers. The insets display the spectral directional emissivity $\mathcal{E}(\mathbf{k})=[\mathcal{E}_{\mathrm{TE}}(\mathbf{k})+\mathcal{E}_{\mathrm{TM}}(\mathbf{k})]/2$ and normal incidence transmission $\mathcal{T}(\theta=0)$ for a cavity where light-matter resonance with a high-order mode at normal incidence is satisfied : $\omega_{\mathrm{cav}}=\omega_{\mathrm{mol}}/3=1000\mathrm{cm}^{-1}/3=333\mathrm{cm}^{-1}$ in (a) and $\omega_{\mathrm{cav}}=\omega_{\mathrm{mol}}/4=2000\mathrm{cm}^{-1}/4=500\mathrm{cm}^{-1}$ in (b). The veritcal red line in the $\mathcal{T}(\theta=0)$ plot and the horizontal red line in the $\mathcal{E}(\mathbf{k})$ plot denote the molecular resonance $\omega_{\mathrm{mol}}$. The cavity+molecule emissivity can be understood as the filtering of molecular emissivity through non-absorbing mirrors in (a) and approximately that of the empty absorbing cavity in (b). In (b), smaller variations in $\mathcal{E}_{\mathrm{tot}}$ are observed as the emissivity is dominated by absorption of CaF$_2$ windows for $\omega<1000\mathrm{cm}^{-1}$ (as seen in the inset).}
\end{figure*}

The microcavity system dissipates energy to surrounding air via: (i) radiation and (ii) convection. We simplify the analysis by assuming local thermal equilibrium between the molecules and cavity, both sharing the same temperature $T_{\mathrm{mol}}$ (this assumption is actually violated in \cite{jarc2023cavity}, as we discuss later) while the temperature of external air equals $T_{\mathrm{out}}$ (Fig. \ref{fig:1}a). The net power lost through radiation follows the Stefan-Boltzmann law, expressed as
\begin{equation}\label{eq:rad}
P_{\mathrm{rad}}=\mathcal{E}_{\mathrm{tot}}\sigma A(T_{\mathrm{mol}}^4-T_{\mathrm{out}}^4),
\end{equation} 
while that through convection is given by 
\begin{equation}\label{eq:conv}
P_{\mathrm{conv}}=\kappa_{\mathrm{conv}}A(T_{\mathrm{mol}}-T_{\mathrm{out}}).    
\end{equation}
Here, $\sigma=\frac{2\pi^5k_B^4}{15c^2h^3}$ denotes the Stefan-Boltzmann constant, $A$ is the surface area of the emitting object, $\mathcal{E}_{\mathrm{tot}}$ is its emissivity, and $\kappa_{\mathrm{conv}}$ is the convective heat transfer coefficient. For an object losing energy to air through free convection, $\kappa_{\mathrm{conv}}$ is $\sim 2.5 \mathrm{Wm}^{-2}\mathrm{K}^{-1}$ \cite{kosky2012exploring}. Radiation and convection contribute to the power lost by a blackbody to air with comparable magnitudes when the temperature difference is $\lesssim 50$K and air is at room temperature; this can be verified using $P_{\mathrm{rad}}/P_{\mathrm{conv}}=\sigma(T_{\mathrm{mol}}^4-T_{\mathrm{out}}^4)/\kappa_{\mathrm{conv}}(T_{\mathrm{mol}}-T_{\mathrm{out}})$ with $T_{\mathrm{out}}=298$K, $\kappa_{\mathrm{conv}}=2.5 \mathrm{Wm}^{-2}\mathrm{K}^{-1}$ and $\sigma=5.67\times 10^{-8}\mathrm{Wm}^{-2}\mathrm{K}^{-4}$. The heat lost by the cavity through convection is readily calculated using Eq. \ref{eq:conv}, while that through radiation is more involved since we need to calculate its emissivity $\mathcal{E}_{\mathrm{tot}}$.

 For typical infrared or UV cavities where the lateral size $R\gg L_z$ (see Fig. \ref{fig:1}a), we can neglect radiation through $A_{||}=2\pi R L_z$ and approximate $A\approx A_{\perp}=2\pi R^2$. The emissivity $\mathcal{E}_{\mathrm{tot}}$ of such a cavity can be computed directly using Maxwell's equations and the fluctuation-dissipation theorem \cite{luo2004thermal,narayanaswamy2004thermal}; however, this is a complex task and a simpler strategy involves obtaining the emissivity indirectly from absorption as Kirchhoff's law dictates that they must be equal \cite{cornelius1999modification}. Following the second approach, we obtain the spectral directional emissivity, $\mathcal{E}_{h}(\mathbf{k})$, using Kirchoff's law $\mathcal{E}_{h}(\mathbf{k})=\mathcal{A}_{h}(\mathbf{k})=1-\mathcal{R}_{h}(\mathbf{k})-\mathcal{T}_{h}(\mathbf{k})$ where we compute the absorption from transfer matrix simulations. Here, $\mathcal{R}_{h}(\mathbf{k})$, $\mathcal{T}_{h}(\mathbf{k})$, and $\mathcal{A}_{h}(\mathbf{k})$ are the reflection, transmission, and absorption, respectively, for incident light with wavevector $\mathbf{k}$ and polarization $h=\mathrm{TE,TM}$. 

For a given $k=|\mathbf{k}|$, integrating $\mathcal{E}_{h}(\mathbf{k})$ over all $\phi\in [0,2\pi]$ and a restricted range of $\theta\in [0,\pi/2]$ and summing the TE/TM contributions gives the spectral hemispherical emissivity $\mathcal{E}(k)$,
\begin{equation}
	\mathcal{E}(k)=\int_0^{\pi/2}d\theta\sin2\theta\frac{1}{(2\pi)}\int_0^{2\pi}d\phi\Big[\frac{1}{2}\sum_{h=\mathrm{TE,TM}}\mathcal{E}_h(\mathbf{k})\Big]
\end{equation}
(Supplementary S1). The total emissivity can then be calculated by integrating $\mathcal{E}(k)$ over all $x\in[0,\infty)$ where $x=\hbar\omega_k/k_BT=\hbar ck/k_BT$,
 \begin{equation}\label{eq:etot}
	\mathcal{E}_{\mathrm{tot}}=\Big(\frac{15}{\pi^4}\Big) \int_0^{\infty}dx\frac{ x^3}{e^{x}-1}\mathcal{E}\Big(x\frac{k_BT}{\hbar c}\Big).
\end{equation}
 Note that $\mathcal{E}(k)\le1$ and when it equals $1$ for all $k$, the total emissivity $\mathcal{E}_{\mathrm{tot}}=1$ and the object becomes a blackbody.

The maximum of the function $\frac{x^3}{e^x-1}$ in Eq. \ref{eq:etot} occurs at $x=2.82$; this condition gives the well-known Wien's displacement law for a blackbody, $\nu_{\mathrm{peak}}=\frac{\omega_{\mathrm{peak}}}{2\pi}=\frac{2.82k_B}{h}T=0.059\mathrm{THz.K}^{-1}T$. The value of $\mathcal{E}(k)$ around $k=2.82\frac{k_BT}{\hbar c}$ dominates the contribution to $\mathcal{E}_{\mathrm{tot}}$. Thus, at room temperature, $\mathcal{E}_{h}(\mathbf{k})$ within the frequency range $200\mathrm{cm}^{-1}<\omega_{\mathbf{k}}<1500\mathrm{cm}^{-1}$ has the largest impact on the total emissivity $\mathcal{E}_{\mathrm{tot}}$. From here on, we consider a temperature of $T=298$K for all emissivity calculations unless specified otherwise.

To explore variations in emissivity with cavity frequency, $\omega_{\mathrm{cav}}$, and its potential impact on temperature-induced modifications in reaction rate constants, we conduct transfer matrix simulations on the structure depicted in Fig. \ref{fig:transfer}. This includes an absorbing molecular layer, spacers, mirrors, and windows of thicknesses $d_{\mathrm{mol}}=3\mu$m, $d_{\mathrm{s}}=(L_z-d_{\mathrm{mol}})/2$, $d_{\mathrm{mirror}}=18$nm, and $d_{\mathrm{w}}=2$mm, respectively, based on characteristic values in experiments \cite{ahn2023modification}, for a range of mirror separations $L_z$. Additionally, their respective refractive indices are taken to be $n_{\mathrm{mol}}(\omega)=\sqrt{\epsilon_{\mathrm{mol}}(\omega)}$,  $n_{\mathrm{s}}=1.43$, $n_{\mathrm{mirror}}(\omega)=\sqrt{\epsilon_{\mathrm{mirror}}(\omega)}$, and $n_{\mathrm{w}}(\omega)$ where the molecular permittivity is modeled using a Lorentzian $\epsilon_{\mathrm{mol}}(\omega)=\epsilon_{\infty}+\frac{f_{\mathrm{mol}}\omega_{\mathrm{p,mol}}^2}{\omega_{\mathrm{mol}}^2-\omega^2-i\Gamma_{\mathrm{mol}}\omega}$ and the mirrors using a Drude permittivity $\epsilon_{\mathrm{mirror}}(\omega)=1-\frac{\omega_{\mathrm{p,mirror}}^2}{\omega^2+i\Gamma_{\mathrm{mirror}}\omega}$. We treat the $2$mm thick windows incoherently and all other layers coherently in our transfer matrix simulations \cite{centurioni2005generalized} (Supplementary S2). Calculations are performed for two parameter sets: (i) with non-absorbing mirrors and windows (ideal system) and (ii) absorbing mirrors and windows (realistic system).

In the first set (Fig. \ref{fig:emissivity}a), the refractive index of the windows is real-valued, $n_{\mathrm{w}}(\omega)=1.43$, and we consider a non-absorbing mirror with parameters $\Gamma_{\mathrm{mirror}}=0$ and $\omega_{\mathrm{p,mirror}}=20000\mathrm{cm}^{-1}$. Given that the refractive indices of the window, mirror, and spacer are all real-valued, they do not absorb, and the molecular layer is solely responsible for all absorption and emission. For the molecular layer, we consider a resonance centered at $\omega_{\mathrm{mol}}=1000\mathrm{cm}^{-1}$ and other parameters $\epsilon_{\infty}=(1.4)^2$, $\omega_{\mathrm{p,mol}}=100\mathrm{cm}^{-1}$, $f_{\mathrm{mol}}=1$ and $\Gamma_{\mathrm{mol}}=200\mathrm{cm}^{-1}$. 

For the second set of parameters (Fig. \ref{fig:emissivity}b), we use $\mathrm{CaF}_{2}$ windows with a complex-valued refractive index $n_{\mathrm{w}}(\omega)$ taken from experiments \cite{li2017new,kelly2017complex}. $\mathrm{CaF}_{2}$ windows are commonly used in vibrational strong coupling chemical reactivity experiments \cite{wiesehan2021negligible,imperatore2021reproducibility,ahn2023modification,vergauwe2019modification}. As 2mm thick $\mathrm{CaF}_{2}$ windows absorb a substantial portion of the radiation when $\omega<1000\mathrm{cm}^{-1}$ (as observed in the inset of Fig. \ref{fig:emissivity}b) dominating the total emissivity, the variation of $\mathcal{E}_{\mathrm{tot}}$ with $\omega_{\mathrm{cav}}$ is smaller than that in Fig. \ref{fig:emissivity}a. On the other hand, 2mm of $\mathrm{ZnSe}$, transparent for $\omega>500\mathrm{cm}^{-1}$ and also previously used for strong coupling experiments \cite{thomas2016ground,thomas2019tilting,imperatore2021reproducibility,hirai2020modulation}, will likely show larger changes in emissivity with cavity length. For the mirror, we adopt Drude parameters of gold, $\hbar\omega_{\mathrm{p,mirror}}=8.5\mathrm{eV}$ and $1/\Gamma_{\mathrm{mirror}}=14$fs \cite{olmon2012optical} and for molecular parameters, we consider a resonance centered at $\omega_{\mathrm{mol}}=2000\mathrm{cm}^{-1}$ with $\epsilon_{\infty}=(1.4)^2$, $\omega_{\mathrm{p,mol}}=100\mathrm{cm}^{-1}$, $f_{\mathrm{mol}}=1$ and $\Gamma_{\mathrm{mol}}=200\mathrm{cm}^{-1}$. 

\begin{figure*}
	\includegraphics[width=\textwidth]{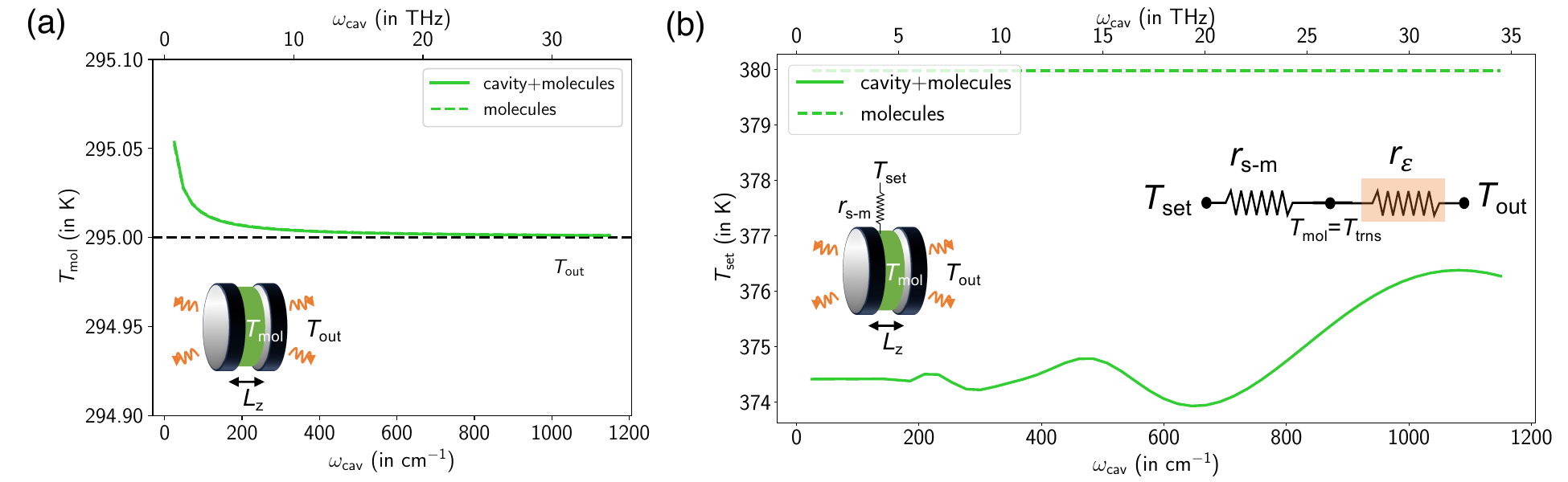}
	\caption{\label{fig:reacttrns} \textbf{Molecular and phase transition temperatures.} (a) The molecular temperature $T_{\mathrm{mol}}$ is plotted against cavity frequency $\omega_{\mathrm{cav}}$ when the temperature of surrounding air is $T_{\mathrm{out}}=295$K (dashed black line) and $k(T_{\mathrm{out}})=5.9\times10^{-6}\mathrm{M}^{-1}s^{-1}$ for the case of bare molecules (dashed green line) and molecules in a cavity (solid green line). The dashed lines are exactly on top of the solid lines and are not clearly visible. (b) In the inset on the left, a schematic illustrates the sample undergoing a phase transition at temperature $T_{\mathrm{mol}}=T_{\mathrm{trns}}$, placed within a cavity and connected to an external heating device at temperature $T_{\mathrm{set}}$ controlling its temperature. A thermal resistance $r_{\mathrm{s-m}}$ exists between the external heating device and sample. The inset on the right depicts the heat flow circuit where $r_{\varepsilon}$ is the thermal resistance between molecules and outside air. The external device temperature $T_{\mathrm{set}}$ required for the phase transition when $T_{\mathrm{out}}=298$K is plotted against the cavity frequency $\omega_{\mathrm{cav}}$ when the sample is placed between non-absorbing windows (dashed green line) and when it is placed inside a non-absorbing cavity (solid green line).}
\end{figure*}

Fig. \ref{fig:emissivity}a-b depict $\mathcal{E}_{\mathrm{tot}}$ as a function of $\omega_{\mathrm{cav}}$ for the two parameter sets mentioned above. These parameters were used to calculate the cavity+molecules (blue line) plots in Fig. \ref{fig:emissivity}a-b. To obtain the plots for molecules (solid orange line), $d_{\mathrm{mirror}}=18$nm was changed to $d_{\mathrm{mirror}}=0$, and for cavity (dashed orange line), $f_{\mathrm{mol}}=1$ was changed to $f_{\mathrm{mol}}=0$. The cavity+molecules plots reveal oscillatory behavior of $\mathcal{E}_{\mathrm{tot}}$ with extrema not necessarily appearing when the resonance condition $\omega_{\mathrm{mol}}=j\omega_{\mathrm{cav}}$ is satisfied (dashed black lines in Fig. \ref{fig:emissivity}); here, $j$ is a positive integer. The reason for oscillations of emissivity with $\omega_{\mathrm{cav}}$ differ between Fig. \ref{fig:emissivity}a and \ref{fig:emissivity}b. In Fig. \ref{fig:emissivity}a, filtering of the emissivity of molecules through the non-absorbing cavity is responsible for this variation as clearly seen from the fact that there is no variation in the bare molecular emissivity (solid orange line) and that the empty cavity has zero emissivity (dashed orange line). In contrast, in Fig. \ref{fig:emissivity}b, the emissivity of the empty cavity (dashed orange line) closely matches that of the cavity+molecule system (blue line), indicating the dominant role played by the absorbing mirrors and windows in determining the emissivity of this structure. Changes in emissivity with $\omega_{\mathrm{cav}}$ are $\sim 0.002 $ in Fig. \ref{fig:emissivity}b which is smaller than $\sim 0.02$ in Fig. \ref{fig:emissivity}a, as it is dominated by the emissivity of the CaF$_2$ windows. The oscillation in emissivity with $\omega_{\mathrm{cav}}$ in Fig. \ref{fig:emissivity}b is due to changes in the modes of the cavity present above the absorption cut-off set by the CaF$_2$ windows ($>1000 \mathrm{cm}^{-1}$, see Fig. \ref{fig:emissivity}b inset). 

\textit{Chemical reaction.--} 
We assume variations in the measured rate constant in vibrational strong coupling experiments, without optical pumping (vibropolaritonic chemistry), stem solely from differences between the outside temperature $T_{\mathrm{out}}$ and the real temperature of molecules $T_{\mathrm{mol}}$. The rate constant for a reaction with activation energy $\mathrm{Ea}$ is $k(T)=k_0e^{-\frac{\mathrm{Ea}}{k_BT}}$. For a bimolecular reaction, $ -C^2\Delta H_{\mathrm{rxn}}k(T_{\mathrm{mol}})V$ is the rate of energy generation inside a volume $V$, where $C$ is the reactant concentration and $\Delta H_{\mathrm{rxn}}$ is the molar enthalpy change. The energy generation rate during the reaction should equal the rate of energy leaving the cavity $P_{\mathrm{rad}}+P_{\mathrm{conv}}$ at steady state (Supplementary S3). This leads to an expression satisfied by $T_{\mathrm{mol}}$,
\begin{equation}\label{eq:react}
	 \begin{aligned}
	 -C^2\Delta H_{\mathrm{rxn}}k(T_{\mathrm{mol}})=\frac{A}{V}\Big[&\sigma \mathcal{E}_{\mathrm{tot}} (T_{\mathrm{mol}}^4-T_{\mathrm{out}}^4)\\
	 &+\kappa_{\mathrm{conv}}(T_{\mathrm{mol}}-T_{\mathrm{out}})\Big]
	 \end{aligned}
\end{equation}
where $V=\pi R^2L_z$ is the volume of the cell. In Eq. \ref{eq:react}, two factors change with $L_z$: firstly, $A/V=2/L_z$, and secondly $\mathcal{E}_{\mathrm{tot}}$ as observed in Fig. \ref{fig:emissivity}.

 To maintain a $10$K temperature difference with external air at $T_{\mathrm{out}}=300$K, we estimate the power that needs to be generated by the reaction to be $1.8\times 10^7 \mathrm{W.m}^{-3}= 4.3 \mathrm{kcal.s}^{-1}\mathrm{L}^{-1}$ for a blackbody in a cavity of size $L_z=10\mu$m using the right-hand side of Eq. \ref{eq:react}. For a typical reaction, $C=1$M, $\Delta H_{\mathrm{rxn}}=-20\mathrm{kcal.mol}^{-1}$ and $k=10^{-5}\mathrm{M}^{-1}\mathrm{s}^{-1}$. Plugging these into the left-hand side of Eq. \ref{eq:react} yields $2\times 10^{-4}\mathrm{kcal.s}^{-1}\mathrm{L}^{-1}$ which is orders of magnitude smaller than the required power of $4.3 \mathrm{kcal.s}^{-1}\mathrm{L}^{-1}$. Therefore, we expect the difference between  $T_{\mathrm{mol}}$ and $T_{\mathrm{out}}$ to be negligible. 
 
 We solve Eq. \ref{eq:react} for $T_{\mathrm{mol}}$ with $\mathcal{E}_{\mathrm{tot}}$ computed using refractive indices as in Fig. \ref{fig:emissivity}b but without spacers ($d_{\mathrm{s}}=0$ and $d_{\mathrm{mol}}=L_z$) (see Fig. S1a), and we take reaction parameters from \cite{ahn2023modification}: $\Delta H =-20.5\mathrm{kcal.mol}^{-1}$, $C=3$M, $k=5.9\times10^{-6}\mathrm{M}^{-1}\mathrm{s}^{-1}$, $T_{\mathrm{out}}=295$K and $\mathrm{Ea}=6.7\mathrm{kcal.mol}^{-1}$. In Fig. \ref{fig:reacttrns}a, the molecular temperature $T_{\mathrm{mol}}$ is plotted against $\omega_{\mathrm{cav}}$ with a dashed green line for the case of bare molecules (structure in Fig. \ref{fig:transfer} with $d_{\mathrm{s}}=0$, $d_{\mathrm{mirror}}=0$ and $d_{\mathrm{mol}}=L_z$) and a solid green line for molecules in a cavity (structure in Fig. \ref{fig:transfer} with $d_{\mathrm{s}}=0$, $d_{\mathrm{mirror}}=18$nm and $d_{\mathrm{mol}}=L_z$). We observe an increase in $T_{\mathrm{mol}}-T_{\mathrm{out}}\sim 0.05$K as $\omega_{\mathrm{cav}}$ decreases ($L_z$ increases; the smallest $\omega_{\mathrm{cav}}$ used in Fig. \ref{fig:reacttrns}a is $25\mathrm{cm}^{-1}$) due to $A/V=2/L_z$ in Eq. \ref{eq:react}. As $L_z$ increases, $V$ increases, leading to an increase in the amount of heat generated by the reaction while the area $A$ through which the heat can escape remains constant, as a result, $T_{\mathrm{mol}}$ increases. In the limit $L_z \to \infty$, we see from Eq. \ref{eq:react} that $T_{\mathrm{mol}}$ approaches infinity. Notably, in Fig. \ref{fig:reacttrns}a, there is no discernible difference between cases with and without the cavity, and the temperature difference between $T_{\mathrm{mol}}$ and $T_{\mathrm{out}}$ is minimal for cavities with $\omega_{\mathrm{cav}} > 25\mathrm{cm}^{-1}$.

\textit{Phase transition.--} When measuring phase transition temperature changes in microcavities, the sample's temperature $T_{\mathrm{mol}}$ is externally adjusted to approach the transition temperature $T_{\mathrm{trns}}$. As $T_{\mathrm{trns}}$ may deviate from $T_{\mathrm{out}}$, there is constant net energy flux between the sample and external air during the measurement. Additionally, the device actively maintaining the sample temperature is at $T_{\mathrm{set}}$ which may not equal $T_{\mathrm{mol}}$ due to potential thermal resistance $r_{\mathrm{s-m}}$ between them, as illustrated in the inset of Fig. \ref{fig:reacttrns}b along with the circuit. As previously assumed, the sample and entire cavity structure share the same temperature $T_{\mathrm{mol}}$. The thermal resistance between the cavity and external air $r_{\varepsilon}$ depends on its emissivity. At steady state, we have 
\begin{equation}			\frac{T_{\mathrm{set}}-T_{\mathrm{mol}}}{r_{\mathrm{s-m}}}=P_{\mathrm{rad}}+P_{\mathrm{conv}}
\end{equation}
which can be rewritten as
\begin{equation}
	\frac{T_{\mathrm{set}}-T_{\mathrm{mol}}}{r_{\mathrm{s-m}}}=\frac{T_{\mathrm{mol}}-T_{\mathrm{out}}}{r_{\varepsilon}}
\end{equation}
where (see Eq. \ref{eq:rad} and \ref{eq:conv})
\begin{equation}
	r_{\varepsilon}=\frac{1}{A\Big[\mathcal{E}_{\mathrm{tot}}\sigma(T_{\mathrm{mol}}^2+T_{\mathrm{out}}^2)(T_{\mathrm{mol}}+T_{\mathrm{out}})+\kappa_{\mathrm{conv}}\Big]}.
\end{equation}
Therefore, $T_{\mathrm{set}}$ required to achieve the transition $T_{\mathrm{mol}}=T_{\mathrm{trns}}$ can vary as $r_{\varepsilon}$ changes. This can lead to experimentally observed cavity-dependent modifications in the external device temperature $T_{\mathrm{set}}$ required for the phase transition as seen in \cite{brawley2023sub} and \cite{jarc2023cavity}. 

In contrast to reactions, where energy generated by the reaction is crucial for maintaining a temperature distinct from the surroundings, systems undergoing phase transitions are actively maintained at a temperature different from surrounding air, $T_{\mathrm{mol}}=T_{\mathrm{trns}}\neq T_{\mathrm{out}}$ through an external device. To demonstrate that modifications in emissivity indeed lead to changes in $T_{\mathrm{set}}$ required for the transition, we consider a transition temperature $T_{\mathrm{trns}}=350$K and a thermal resistance of $r_{\mathrm{s-m}}=229\mathrm{K.W}^{-1}$. We choose this value of $r_{\mathrm{s-m}}$ by applying the condition $T_{\mathrm{set}}-T_{\mathrm{trns}}=30$K in the absence of mirrors ($\mathcal{E}_{\mathrm{tot}}=0.19$ - solid orange line in Fig. S1b) for the structure in Fig. \ref{fig:transfer} when $T_{\mathrm{out}}=298$K and the area $A=\pi R^2$ with $R=1$cm (\cite{jarc2023cavity} has a $35-40$K difference between $T_{\mathrm{set}}$ and $T_{\mathrm{trns}}$ in free space). We use the same parameters for the emissivity $\mathcal{E}_{\mathrm{tot}}$ calculation as in Fig. \ref{fig:emissivity}a but with temperature $350$K (see Fig. S1b). We plot the results in Fig. \ref{fig:reacttrns}b, revealing that the temperature follows the same trend as the emissivity. Note that these results are for the ideal system of non-absorbing windows and mirrors. For the realistic system with CaF$_2$ windows and gold mirrors, we find negligible changes $\sim 0.5$K in $T_{\mathrm{set}}$ with $\omega_{\mathrm{cav}}$.

\begin{figure}
	\includegraphics[width=\columnwidth]{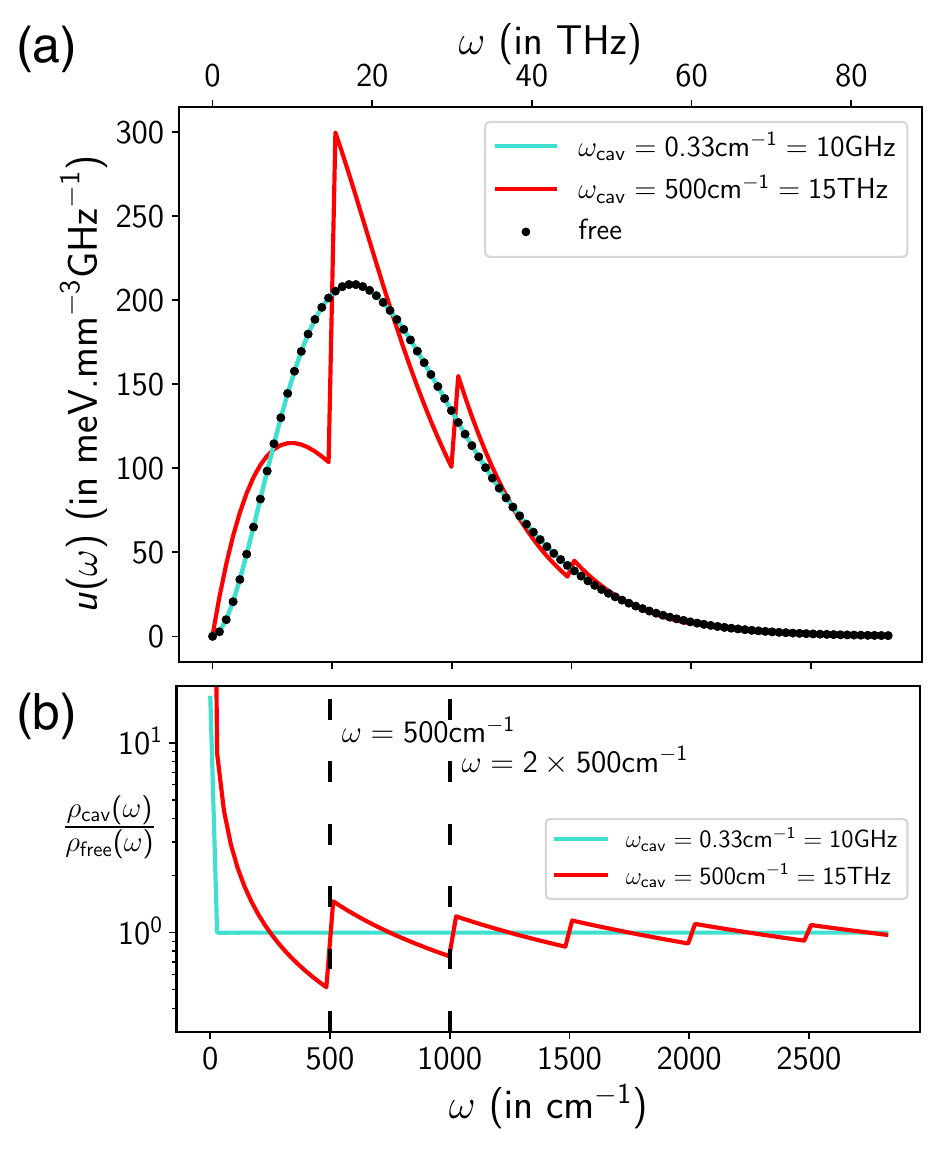}
	\caption{\label{fig:blackbody} \textbf{Thermal energy density and photon density of states (DoS) in a cavity.} (a) Thermal energy density inside a cavity of frequency $\omega_{\mathrm{cav}}=500\mathrm{cm}^{-1}$ and $\omega_{\mathrm{cav}}=0.33\mathrm{cm}^{-1}$ are plotted in red and cyan, respectively. The black dots represent thermal energy density in free space. (b) Ratio of photon DoS inside a cavity of frequency $\omega_{\mathrm{cav}}=500\mathrm{cm}^{-1}$, $\rho_{\mathrm{cav}}(\omega)$, and in free space, $\rho_{\mathrm{free}}(\omega)$.}
\end{figure}

\textit{Blackbody radiation.--} We showed that changes in chemical reaction rates cannot be explained through differences in temperature between molecules and external air; however, even at the same temperature, the thermal energy density inside a cavity is different from that in free space and may lead to changes in reactivity. The thermal energy density, $u_x(\omega)$, is given by the product of mean energy in a mode $\Theta(T,\omega)=\hbar\omega\langle n\rangle=\frac{\hbar\omega}{e^{\hbar\omega/k_{\mathrm{B}}T}-1}$ and the photon density of states (DoS) $\rho_x(\omega)$ where $x=\mathrm{cav,free}$ \cite{reiser2013geometric}. Here, we ignore the zero-point contribution $\hbar\omega/2$ to the mean energy $\Theta(T,\omega)$ \cite{reggiani2022revisiting}. The energy density inside a cavity of frequency $\omega_{\mathrm{cav}}$ is 
\begin{equation}
	\begin{aligned}
		u_{\mathrm{cav}}(\omega)
		=&\frac{\omega n^3\omega_{\mathrm{cav}}}{\pi^2 c^3}\Bigg(\frac{1}{2}+\Big\lfloor\frac{\omega}{\omega_{\mathrm{cav}}}\Big\rfloor\Bigg)\Bigg(\frac{\hbar\omega}{e^{\hbar\omega/k_{\mathrm{B}}T}-1}\Bigg).
	\end{aligned}
\end{equation}
 In the limit $L_z\to \infty$ or equivalently $\omega_{\mathrm{cav}}\to 0$, this reduces to free space blackbody radiation,
\begin{equation}
	\begin{aligned}
		u_{\mathrm{free}}(\omega)=\frac{\omega^2n^3}{\pi^2 c^3}\Bigg(\frac{\hbar\omega}{e^{\hbar\omega/k_{\mathrm{B}}T}-1}\Bigg).
	\end{aligned}
\end{equation}
We plot $u_{\mathrm{cav}}(\omega)$ for two cavity frequencies $\omega_{\mathrm{cav}}=500\mathrm{cm}^{-1},0.33\mathrm{cm}^{-1}$ at $T=298$K using solid lines and $u_{\mathrm{free}}(\omega)$ at the same temperature using black circles in Fig. \ref{fig:blackbody}a. 

The ratio of the spectral energy densities in a cavity and in free space equals $\frac{u_{\mathrm{cav}}(\omega)}{u_{\mathrm{free}}(\omega)}=\frac{\rho_{\mathrm{cav}}(\omega)}{\rho_{\mathrm{free}}(\omega)}$ where,
\begin{equation}\label{eq:ratio}
	\frac{\rho_{\mathrm{cav}}(\omega)}{\rho_{\mathrm{free}}(\omega)}=\frac{\omega_{\mathrm{cav}}}{\omega}\Big\lfloor\frac{\omega}{\omega_{\mathrm{cav}}}\Big\rfloor+\frac{\omega_{\mathrm{cav}}}{2\omega}.
\end{equation}
From Eq. \ref{eq:ratio} (plotted in Fig. \ref{fig:blackbody}b), we see in the limit $\omega\to 0$ that $\frac{u_{\mathrm{cav}}(\omega)}{u_{\mathrm{free}}(\omega)}\to \infty$. However, since $\lim_{\omega\to 0}u_{\mathrm{cav}}(\omega)=0$ and $\lim_{\omega\to 0}u_{\mathrm{free}}(\omega)=0$, their difference goes to $0$ when $\omega\to 0$ as seen in Fig. \ref{fig:blackbody}a.

We expect the energy density within a cavity to differ from that in free space appreciably only when $\omega_{\mathrm{peak}}/\omega_{\mathrm{cav}}\lesssim 1$, \textit{i.e.}, 
\begin{equation}\label{eq:condition}
	2.82k_BT/\hbar\omega_{\mathrm{cav}}\lesssim 1,
\end{equation}
as seen from Fig. \ref{fig:blackbody}a. A similar result was noted in the context of thermal emission from photonic band gap materials \cite{cornelius1999modification}. 
Interestingly, this condition is satisfied by cavities used in most experiments of modified chemical reactivity under collective vibrational strong coupling \cite{thomas2019tilting,ahn2023modification}. This difference in energy density may have an effect on chemical reactivity as it shows sharp changes at the normal incidence ($k_{||}=0$) cavity frequency $\omega_{\mathrm{cav}}$ (Fig. \ref{fig:blackbody}a). This is an interesting direction that we will explore in more detail in future work. In previous work, it has been shown that the photon DoS is different inside a cavity \cite{vurgaftman2022comparative} and can potentially modify chemical reactivity even in the dark and may explain resonance effects in polariton chemistry \cite{ying2023resonance}. 

Notice, on the other hand, how the thermal spectral energy density within a $\omega_{\mathrm{cav}}=10\mathrm{GHz}$ cavity -- similar to those used by \cite{jarc2023cavity} in their phase transition experiments -- is close to that in free space. Furthermore, our theory for modification of observed phase transition temperatures shows changes for IR cavities but does not explain their experiment as the temperature change (Fig. \ref{fig:reacttrns}b) and emissivity (Fig. \ref{fig:emissivity}) become a constant for low frequency cavities $\omega_{\mathrm{cav}}<5\mathrm{THz}$ due to the condition in Eq. \ref{eq:condition}. However, we have assumed that the temperature of the entire cavity structure is a constant while in their experiments, the mirror and sample have different temperatures and when they vary the mirror temperature, the observed phase transition temperature changes. Our emissivity calculations can be generalized to structures with non-uniform temperature \cite{wang2011direct} and the distance-dependent changes in the rate of radiative heat transfer between the mirror and sample might explain their work. 

As pointed out in \cite{chiriaco2023thermal}, the theoretical framework presented in \cite{jarc2023cavity} does not accurately account for the photon DoS in a Fabry-Perot cavity. Chiriac{\`o} \cite{chiriaco2023thermal} explains the experiments in \cite{jarc2023cavity} using a thermal Purcell effect, arising from the difference in photon DoS between the cavity and free space. They expect the heat transfer rate of a sample inside a Fabry-Perot cavity with $\omega_{\mathrm{cav}}\sim 10$GHz to be the same as in free space because the photon DoS of such a cavity is nearly identical to that in free space for the frequency range over which the blackbody energy density is substantial. Consequently, they claim that a cavity with spherical mirrors is required to explain the observations in \cite{jarc2023cavity}. In their model, the photon DoS along with the temperature difference between the sample and cavity photons determines the net heat flux between them and plays a crucial role. 
Their conclusions hold if the photon modes have a well-defined temperature; however, in the experiments of \cite{jarc2023cavity}, where mirrors and sample are at different temperatures, the cavity photons do not have a well-defined temperature due to interactions with both. In our calculations, we do not need to define a temperature for the photon modes; the net heat flux can be computed as the rate of radiative heat transfer between the sample and mirrors when altering the distance between the mirrors. This may explain results of \cite{jarc2023cavity}, even in a Fabry-Perot cavity, as discussed in the previous paragraph.

Additionally, the lowest TM$_0$ mode (see Fig. 1) was not included in \cite{chiriaco2023thermal} while calculating the photon DoS of a Fabry-Perot cavity, leading to the incorrect conclusion that the photon DoS inside a Fabry-Perot cavity is always smaller than that in free space for all frequencies (however, this does not affect the main conclusion of their paper, as $u(\omega)$ in a Fabry-Perot cavity with $\omega_{\mathrm{cav}}\sim 10$GHz is very close to that in free space; see Fig. \ref{fig:blackbody}a).

\section*{Conclusions}
In summary, we theoretically demonstrate that temperature differences between reacting molecules in a cavity and external air are negligible -- ruling out the possibility that collective VSC chemical reaction rate modifications can be explained purely through increase/decrease in the temperature within a cavity from the heat generated/absorbed during a reaction. In contrast, our calculations reveal that cavities indeed exert a notable thermal influence on observed phase transition temperatures if cavity windows are non-absorbing (ideal system). In this case, the material's temperature is actively controlled by a heating/cooling source, and significant temperature gradients exist between the material and surrounding air, as observed in recent experiments \cite{brawley2023sub} and \cite{jarc2023cavity}. Additionally, we show that the blackbody spectrum inside Fabry-Perot cavities differs from free space when they satisfy $2.82k_BT\lesssim \hbar\omega_{\mathrm{cav}}$. As this condition is typically satisfied by the IR cavities used for VSC experiments and there are discontinuities in the blackbody spectrum at multiples of the cavity frequency, differences in the blackbody spectrum can potentially explain the resonance effect observed in reaction rates. This will be explored in future work.


\begin{acknowledgements}
 S.P.S. thanks Gerrit Groenhof for useful discussions. S.P.S. and J.Y.Z acknowledge funding support from the W. M. Keck Foundation. 
 \end{acknowledgements}

\section*{Code availability}
 The code used in this work is available at \textcolor{blue}{https://github.com/SindhanaPS/blackbody\_and\_thermal\_effects\_polaritons}.

\end{document}


\title{Supplementary information: Blackbody radiation and thermal effects on chemical reactions and phase transitions in cavities}
	
	\author{Sindhana Pannir-Sivajothi}
	\affiliation{Department of Chemistry and Biochemistry, University of California San Diego, La Jolla, California 92093, USA}
	\author{Joel Yuen-Zhou}
	\email{joelyuen@ucsd.edu}
	\affiliation{Department of Chemistry and Biochemistry, University of California San Diego, La Jolla, California 92093, USA}
	\maketitle

\section{Emissivity}
The right hand side of Eq. \ref{eq:emissivity} calculates the flux of energy through an area defined by $\mathbf{\hat{z}}$, the unit vector perpendicular to the mirror, by integrating over $\mathbf{k}$ restricted to the hemisphere $0 \le \theta \le \pi/2$ to ensure that we only account for outgoing radiation, 
  \begin{equation}\label{eq:emissivity}
	\begin{aligned}
		\mathcal{E}_{\mathrm{tot}}\sigma T^4=&\sum_{h=\mathrm{TE,TM}}\int_{\mathrm{Restricted}} d\mathbf{k} \mathbf{s}_h(\mathbf{k}).\mathbf{\hat{z}}\\
		=&\sum_{h=\mathrm{TE,TM}}\int_{\mathrm{Restricted}} d\mathbf{k} \frac{c}{(2\pi)^3}\mathcal{E}_h(\mathbf{k})\Theta(T,\omega_{\mathbf{k}})\mathbf{\hat{k}}.\mathbf{\hat{z}}\\
		=&\sum_{h=\mathrm{TE,TM}}\int_{\mathrm{Restricted}} d\mathbf{k} \frac{c}{(2\pi)^3}\mathcal{E}_h(\mathbf{k})\Theta(T,\omega_{\mathbf{k}})\cos\theta\\
		=&\frac{c}{(2\pi)^3}\sum_{h=\mathrm{TE,TM}} \int_0^{\infty}dkk^2\Theta(T,\omega_{k})\int_0^{\pi/2}d\theta\sin\theta\cos\theta\int_0^{2\pi}d\phi\mathcal{E}_h(k,\theta,\phi) \\
		=&\frac{c}{(2\pi)^2} \int_0^{\infty}dkk^2\Theta(T,\omega_{k})\Bigg\{\int_0^{\pi/2}d\theta\sin2\theta\frac{1}{(2\pi)}\int_0^{2\pi}d\phi\Big[\frac{1}{2}\sum_{h=\mathrm{TE,TM}}\mathcal{E}_h(k,\theta,\phi)\Big]\Bigg\} \\
		=&\frac{c}{(2\pi)^2} \int_0^{\infty}dkk^2\Theta(T,\omega_{k})\mathcal{E}(k).
	\end{aligned}
\end{equation}
Here, $\Theta(T,\omega)=\hbar\omega\langle n\rangle=\frac{\hbar\omega}{e^{\hbar\omega/k_{\mathrm{B}}T}-1}$ is the mean energy in a mode with frequency $\omega$ at temperature $T$, $k=|\mathbf{k}|$, $\omega_{\mathbf{k}}=\omega_{k}=c|\mathbf{k}|$, and $c$ is the speed of light; $\Theta(T,\omega)$ does not include zero-point energy.
Upon simplification, Eq. \ref{eq:emissivity} gives us 
\begin{equation}
	\begin{aligned}
			\mathcal{E}_{\mathrm{tot}}\sigma T^4=&\frac{c}{(2\pi)^2}k_BT\Big(\frac{k_BT}{\hbar c}\Big)^3 \int_0^{\infty}dxx^2\frac{ x}{e^{x}-1}\mathcal{E}\Big(x\frac{k_BT}{\hbar c}\Big)\\
		=&\frac{c}{(2\pi)^2}k_BT\Big(\frac{k_BT}{\hbar c}\Big)^3 \int_0^{\infty}dx\frac{ x^3}{e^{x}-1}\mathcal{E}\Big(x\frac{k_BT}{\hbar c}\Big)\\
		=&\sigma T^4\Bigg[\Big(\frac{15}{\pi^4}\Big) \int_0^{\infty}dx\frac{ x^3}{e^{x}-1}\mathcal{E}\Big(x\frac{k_BT}{\hbar c}\Big)\Bigg]
	\end{aligned}
\end{equation} 
where $x=\hbar\omega_k/k_BT$.
The total emissivity is
\begin{equation}
	\mathcal{E}_{\mathrm{tot}}=\Big(\frac{15}{\pi^4}\Big) \int_0^{\infty}dx\frac{ x^3}{e^{x}-1}\mathcal{E}\Big(x\frac{k_BT}{\hbar c}\Big).
\end{equation}
	\section{Transfer matrix simulation}
      A transfer matrix $M$ relates the coefficients of the electric field on one side of a structure to the other side \cite{cornelius1999modification},
     \begin{equation}
     	\begin{bmatrix}
     		1\\
     		r
     	\end{bmatrix}
     	=\begin{bmatrix}
     		M_{11} & M_{12} \\
     		M_{21} & M_{22}
     	\end{bmatrix}
     	\begin{bmatrix}
     		t \\
     		0
     	\end{bmatrix}.
     \end{equation}
     From this matrix, we can obtain the reflection $\mathcal{R}$ and transmission $\mathcal{T}$,
     \begin{subequations}
     	\begin{align}
     		\mathcal{T}=& |t|^2=\frac{1}{|M_{11}|^2},\\
     		\mathcal{R}=& |r|^2=\Big|\frac{M_{21}}{M_{11}}\Big|^2.
     	\end{align}
     \end{subequations}

  Consider a structure consisting of $N$ slabs where each slab is labeled with an integer $j$ and has refractive index $n_j(\omega)$. The transfer matrix when light with wavevector $\mathbf{k}$ and polarization $h=\mathrm{TE,TM}$ is incident on such a structure can be calculated using: (i) interface matrices $\Delta_{h,i,j}(\mathbf{k})$ that relate the electric field on the two sides of an interface, where $i$ and $j$ are the indices that label the slabs on either side of the interface and (ii) propagation matrices $\Pi_{j}(\mathbf{k})$ that relate the electric field on one end of a slab $j$ to the other end. The total transfer matrix for $N$ slabs with a medium of index $n_0$ on both ends is $M_{h}(\mathbf{k})=\Big[\prod_{j=0}^{N-1}\Delta_{h,j,j+1}(\mathbf{k})\Pi_{j+1}(\mathbf{k})\Big]\Delta_{h,N,0}(\mathbf{k})$.
  
  The interface matrix,
 \begin{equation}
 	\begin{aligned}
 		\Delta_{h,ij}(\mathbf{k}) =&
 		\begin{bmatrix}
 			\delta^+_{h,ij}(\mathbf{k}) &\delta^-_{h,ij}(\mathbf{k})  \\
 			\delta^-_{h,ij}(\mathbf{k})  & \delta^+_{h,ij}(\mathbf{k})
 		\end{bmatrix}
 	\end{aligned}
 \end{equation}
 and
 \begin{subequations}\label{eq:Fresnel}
 	\begin{align}
 		\delta^{\pm}_{\mathrm{TE},ij}(\mathbf{k})=&\frac{1}{2}\Big(1\pm\frac{n_j(\omega_{\mathbf{k}})\cos\theta_j}{n_i(\omega_{\mathbf{k}})\cos\theta_i}\Big),\\
 		\delta^{\pm}_{\mathrm{TM},ij}(\mathbf{k})=&\frac{1}{2}\Big(\frac{\cos\theta_j}{\cos\theta_i}\pm\frac{n_j(\omega_{\mathbf{k}})}{n_i(\omega_{\mathbf{k}})}\Big),
 	\end{align}
 \end{subequations}
 where $\theta_j$ is the angle of incidence in each slab and Eq. \ref{eq:Fresnel} are determined from Fresnel equations \cite{zangwill2013modern} (In Eq. 27b of \cite{cornelius1999modification}, there is a mistake in $\Delta_{\mathrm{TM},ij}(\mathbf{k})$ ).
We can get $\cos \theta_j$ using 
\begin{equation}
	\cos\theta_j=\pm \frac{1}{n_j(\omega_{\mathbf{k}})}\sqrt{n_j(\omega_{\mathbf{k}})^2-(n_0(\mathbf{k})\sin\theta_0)^2}
\end{equation}
because $n_0\sin\theta_0=n_{j}\sin\theta_{j}$ (Snell's law) for all the layers. Here, $\theta_j$ can be complex as $n_j$ may be complex. We need to choose the sign of $\cos\theta_j$ such that $\Im(n_j\cos\theta_j)>0$ to ensure that $\mathcal{T}\le 1$ \cite{byrnes2016multilayer}. The propagation matrix for layer $j$ of thickness $L_j$ is \cite{cornelius1999modification, jung2011optical}
\begin{equation}
	\Pi_{j}(\mathbf{k})=\begin{bmatrix}
		e^{-in_j(\omega_{\mathbf{k}})|\mathbf{k}|L_j\cos \theta_j} & 0 \\
		0 & e^{in_j(\omega_{\mathbf{k}})|\mathbf{k}|L_j\cos \theta_j}
	\end{bmatrix}.
\end{equation}

 For the CaF$_2$ window mentioned in the main text, we use thickness $d_{\mathrm{w}}=2\mathrm{mm}$ and complex refractive index from \cite{kelly2017complex} for $\omega=0-900\mathrm{cm}^{-1}$ and from \cite{li2017new} for $\omega>900\mathrm{cm}^{-1}$. We treat the thick CaF$_2$ layers incoherently and treat all other layers coherently within the transfer matrix formalism \cite{centurioni2005generalized}. So far, the transfer matrices mentioned were for the coherent case. The incoherent transfer matrix formalism relates the electric field amplitude square $|E|^2$ instead of $E$. For a coherent-incoherent multilayer, the layer of coherent slabs would give a regular transfer matrix $M$ which will become an interface matrix in the incoherent formalism,
\begin{equation}\label{eq:incohtm}
	\bar{I}=	\begin{bmatrix}
	|M_{11}|^2 & -|M_{12}|^2 \\
		|M_{21}|^2 & \frac{|\mathrm{det} \mathbf{M}|^2-|M_{12}M_{21}|^2}{|M_{11}|^2}
	\end{bmatrix}.
\end{equation}
To find the incoherent transfer matrix at the interface between two incoherent layers or to find the incoherent propagation transfer matrix, we take $M=\Delta_{h,ij}(\mathbf{k})$ or $M=\Pi_{j}(\mathbf{k})$, respectively, and use Eq. \ref{eq:incohtm} to compute the corresponding incoherent transfer matrices. The total incoherent transfer matrix can then by obtained by multiplying the incoherent transfer matrices for propagation and interfaces similar to the coherent transfer matrix case.

\section{Reaction rate and heat transfer coupled equations}
In the main text, we consider a steady-state heat flow equation where the heat generated during a reaction leaves the system through convection and radiation (see Eq. 6). In the present section, we justify this steady state approximation; it is valid when the rate at which the concentration of the reactant is changing is much slower than the time-scale associated with the heat flow dynamics. To quantify this, let's consider a first-order chemical reaction and the corresponding heat flow equation
\begin{subequations}
    \begin{align}
        \frac{dC}{dt}=&-Ck(T_{\mathrm{mol}}),\\
        VC_{\mathrm{solv}}c_{\mathrm{solv}}\frac{dT_{\mathrm{mol}}}{dt}=&-VCk(T_{\mathrm{mol}})\Delta H_{\mathrm{rxn}}-(P_{\mathrm{rad}}+P_{\mathrm{conv}}),
    \end{align}
\end{subequations}
where $C$ is the concentration of the reactant, $k(T)$ is the rate constant at temperature $T$, $C_{\mathrm{solv}}$ is the concentration of the solvent, $V$ is the volume of the reaction vessel and $c_{\mathrm{solv}}$ is the molar heat capacity of the solvent. Defining dimensionless variables $x=C/C_0$ and $y=(T_{\mathrm{mol}}-T_{\mathrm{out}})/T_{\mathrm{out}}$ and rewriting the equations, we get a pair of coupled differential equations
\begin{subequations}\label{eq:couplexy}
    \begin{align}
        \frac{dx}{dt}=&-xk(y),\\
        \frac{dy}{dt}=&-xk(y)\frac{C_0\Delta H_{\mathrm{rxn}}}{C_{\mathrm{solv}}c_{\mathrm{solv}}T_{\mathrm{out}}}-f(y)y-\frac{\kappa_{\mathrm{conv}}A}{C_{\mathrm{solv}}c_{\mathrm{solv}}V}y,
    \end{align}
\end{subequations}
 where $C_0$ is the initial $t=0$ concentration of the reactant and
\begin{subequations}
\begin{align}
    k(y)=&k_0\exp\Big(\frac{-\mathrm{Ea}}{k_BT_{\mathrm{out}}(1+y)}\Big),\\
    f(y)=&\frac{\mathcal{E}_{\mathrm{tot}}\sigma AT_{\mathrm{out}}^3}{C_{\mathrm{solv}}c_{\mathrm{solv}}V} \Big[(1+y)^2+1\Big]\Big[(1+y)+1\Big].
\end{align} 
\end{subequations}
The set of coupled differential equations Eq. \ref{eq:couplexy} have equilibrium solutions $x=0$ and $y=0$. 

To identify the timescales associated with the dynamics of $x$ and $y$, we substitute $y=y_0$ in Eq. \ref{eq:couplexy}a and find $\tau_x=1/k(y_0)$ and when $y_0\ll 1$, we get $\tau_x=(k_0e^{-\mathrm{Ea}/k_BT_{\mathrm{out}}})^{-1}$ \cite{eilertsen2020quasi}. Similarly, we substitute $x=x_0$ in Eq. \ref{eq:couplexy}b and find the timescale 
\begin{equation}
    \tau_y=\Bigg(-x_0k_0e^{-\frac{\mathrm{Ea}}{k_BT_{\mathrm{out}}}}\Big(\frac{C_0|\Delta H_{\mathrm{rxn}}|}{C_{\mathrm{solv}}c_{\mathrm{solv}}T_{\mathrm{out}}}\Big)\frac{\mathrm{Ea}}{k_BT_{\mathrm{out}}}+\frac{4\mathcal{E}_{\mathrm{tot}}\sigma AT_{\mathrm{out}}^3}{C_{\mathrm{solv}}c_{\mathrm{solv}}V}+\frac{\kappa_{\mathrm{conv}}A}{C_{\mathrm{solv}}c_{\mathrm{solv}}V}\Bigg)^{-1}
\end{equation}
under the condition that $y\ll 1$ for all $t$ by considering only $O(y)$ terms in Eq. \ref{eq:couplexy}b,
\begin{equation}
    \frac{dy}{dt}=-x_0k_0e^{-\frac{\mathrm{Ea}}{k_BT_{\mathrm{out}}}}\Big(\frac{C_0\Delta H_{\mathrm{rxn}}}{C_{\mathrm{solv}}c_{\mathrm{solv}}T_{\mathrm{out}}}\Big)\Big(1+\frac{\mathrm{Ea}}{k_BT_{\mathrm{out}}}y\Big)-\frac{4\mathcal{E}_{\mathrm{tot}}\sigma AT_{\mathrm{out}}^3}{C_{\mathrm{solv}}c_{\mathrm{solv}}V}y-\frac{\kappa_{\mathrm{conv}}A}{C_{\mathrm{solv}}c_{\mathrm{solv}}V}y
\end{equation}
as $k(y)=k_0\exp(-\mathrm{Ea}/k_BT_{\mathrm{out}})(1+\frac{\mathrm{Ea}}{k_BT_{\mathrm{out}}}y)+O(y^2)$ and $f(y)y=\frac{4\mathcal{E}_{\mathrm{tot}}\sigma AT_{\mathrm{out}}^3}{C_{\mathrm{solv}}c_{\mathrm{solv}}V}y+O(y^2)$.

Therefore, under the condition $y(t)\ll 1$ at all $t$, when $\tau_y\ll \tau_x$, we can apply the quasi steady-state approximation and solve for $y$ for a given $x$ by setting Eq. \ref{eq:couplexy}b to zero \cite{eilertsen2020quasi}. The maximum value of $\tau_y$ occurs when $x_0$ is maximum and this is $x_0=1$. Considering a reaction with $k_0=0.54\mathrm{s}^{-1}$, $\Delta H_{\mathrm{rxn}}=-20.5 \mathrm{kcal.mol}^{-1}$, $\mathrm{Ea}=6.7\mathrm{kcal.mol}^{-1}$ (similar to \cite{ahn2023modification} but converting to first-order reaction parameters by taking the prefactor in the rate constant and multiplying it by $1\mathrm{M}$ to get $k_0$), $C_0=1$M, with tetrahydrofuran (THF) as the solvent $C_{\mathrm{solv}}=0.88\mathrm{g.mL}^{-1}/72\mathrm{g.mol}^{-1}=12\mathrm{M}$, $c_{\mathrm{solv}}=124 \mathrm{J.mol}^{-1}\mathrm{K}^{-1}$, and $T_{\mathrm{out}}=295$K, we find $\tau_x=1.7\times10^5 \mathrm{s}$ and $\tau_y=0.9$s. Clearly, $\tau_y\ll \tau_x$ and we can make the quasi steady-state approximation.

The main idea is -- as $C$ varies slowly compared to $T_{\mathrm{mol}}$, we can fix $C$ and set Eq. \ref{eq:couplexy}b to zero (quasi steady-state approximation) and solve for $T_{\mathrm{mol}}$, so for each $C$ we get the corresponding $T_{\mathrm{mol}}$. Then, we can plug this expression $T_{\mathrm{mol}}(C)$ into Eq. \ref{eq:couplexy}a and solve for $C$ as a function of $t$. In the main text, we want to find $T_{\mathrm{mol}}$ at $t=0$, so we set $C=C_0$ and determine $T_{\mathrm{mol}}$ by setting an equation similar to Eq. \ref{eq:couplexy}b (will be slightly different as we consider a second-order reaction in the main text) to zero.

\begin{figure*}
	\includegraphics[width=\textwidth]{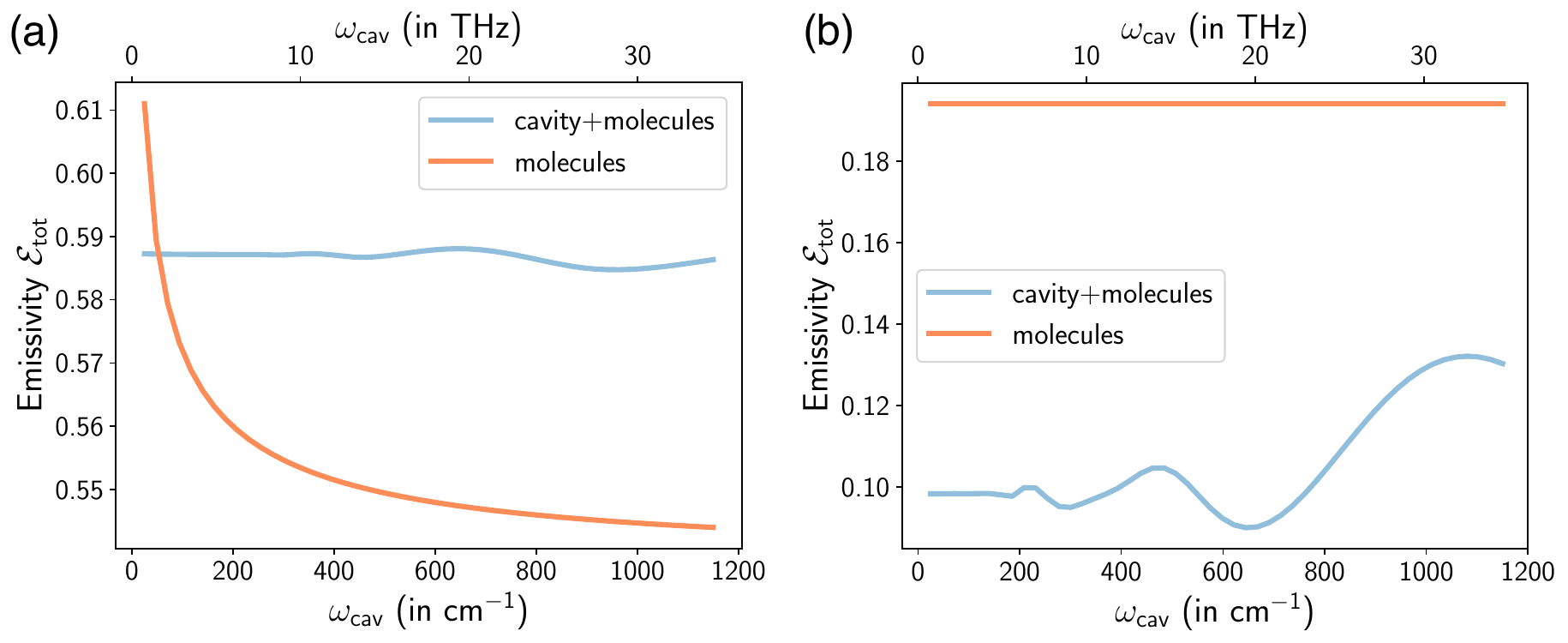}
	\caption{\label{fig:fig4emissivity} \textbf{Emissivity used in Fig. 4.} The emissivity $\mathcal{E}_{\mathrm{tot}}$ used for the calculation in (a) Fig. 4a and (b) Fig. 4b is plotted against cavity frequency $\omega_{\mathrm{cav}}=c\pi/nL_z$.}
\end{figure*}